\documentclass[review]{elsarticle}

\usepackage{lineno,hyperref}
\modulolinenumbers[5]

\journal{Journal of \LaTeX\ Templates}

\begin{document}

\begin{frontmatter}

\title{Gravitational wave background discovered by NANOGrav as evidence of a cyclic universe}

\author{Gorkavyi N.\fnref{mymainaddress}}



\address[mymainaddress]{Science Systems and Applications, Inc., 10210 Greenbelt Road, Lanham, MD 20706, US}

\begin{abstract}
The consortium NANOGrav discovered the isotropic gravitational wave background 
(GWB) with an amplitude of $h\sim 10^{-15}$ and a frequency of $f\sim 10^{-8}$ Hz 
using observations of millisecond pulsars. We hypothesize that the GWB is 
relic radiation left over from the merging stellar mass black holes (SBHs) 
during Big Crunch at $z\sim 10^{10}$. The relic gravitational waves are similar to 
the gravitational waves with $f\sim 10^2$ Hz discovered by LIGO in 2015, taking into 
account a decrease in frequency by a factor $\sim 10^{10}$ due to the expansion of the Universe. 
We take as a basis the observed spectrum of 139 SBHs, discovered by the LIGO observatory.
Our model explains well all the observed features of the GWB. Unlike all other GWB models, 
our model predicts a sharp decrease in the GWB amplitude at frequencies $f>3.5*10^{-8}$ Hz, 
reflecting the deficit of SBHs with masses $<4M_\odot$. The SBMH mergers at Big Crunch 
should generate yet undiscovered GWB with a frequency of  $f\sim 10^{-(14-17)}$ Hz.
\end{abstract}

\begin{keyword}
\texttt Gravitational waves\sep NANOGrav\sep black holes\sep oscillating Universe
\MSC[2010] 83-57\sep  83-35
\end{keyword}

\end{frontmatter}


\section*{Highlights}
\begin{itemize}
\item Gravitational Wave Background (GWB) is relic radiation left over Big Crunch at $z\sim 10^{10}$
\item The relic gravitational waves are similar to the gravitational waves discovered by LIGO
\item Decrease in $10^{10}$ times the frequency of GWB is associated with the expansion of the Universe. 
\item The model predicts a sharp decrease in the GWB amplitude at frequencies $f>3.5*10^{-8}$ Hz,
\end{itemize}
\section{Introduction}
Isotropic CMB radiation discovered by \cite{Penzias} is relic radiation of a compressed hot 
Universe, which decreased its temperature during its expansion \cite{Gamov, Dicke, Mather}.
The relic gravitational waves (GW) were discussed in many papers 
(see, for example, \cite{Zeldovich, Grish}). The most powerful known sources of gravitational 
radiation are merging black holes (BH). In 2015, the LIGO group discovered 
gravitational waves of $f\sim 10^{2}$ Hz caused by the merging of black holes with 
stellar masses (SBH up to 100 solar masses) \cite{Abbot16}. Gravitational radiation from 
modern binary SMBH can generate gravitational waves with a length of about a 
light year, which can be detected by variations in the radio pulses from 
millisecond pulsars  \cite{Sazhin, Detweiler, Jaffe}. Recently, the NANOGrav consortium announced 
the possible discovery of an isotropic stochastic background of gravitational 
waves with a frequency $f\sim 2.5*10^{-9}\div1.2*10^{-8}$ Hz (or $2.5\div12$ nHz) 
and an amplitude of $h\sim 10^{-15}$ \cite{Arz}. It is assumed that these waves are generated by the current 
binary SMBH \cite{Jaffe}, although many other hypotheses of origin the GWB have been put 
forward: from collapsing fluctuations in the early stages of the Big Bang \cite{Nguyen, DeLuca}
to the break of a cosmic string \cite{Buchmuller}. 

Our work analyzes the formation of the GWB in a cyclic Universe. 
The model of an oscillating Universe was popular until the 1980s \cite{Gamov, Dicke, 
Zeldovich, Peebles} and now there is a growing interest in this model \cite{Gurzadyan, Poplawski, Valent, GorkavyiD}. 
The classical cyclic cosmology included a mechanism of the periodical transformation of chemical 
elements: the effective photodissociation of heavy atomic nuclei into individual baryons 
began when the temperature of the electromagnetic relic radiation in the compressed 
Universe reached $\sim 3*10^{10}K$. This is how hydrogen was created to form stars in a new cycle.
The temperature $3*10^{10}K$ is reached at $z\sim1.6*10^{10}$ corresponds to the radius of 
the Universe $\approx3$ ly \cite{Weinberg}. 

In \cite{GorkavyiD, GorkavyiB, GorkavyiC}, a cyclic model of the Universe with a large number of 
black holes has been analyzed (see Fig. 1). Obviously, when the Universe collapses, a significant part of the 
gravitational mass of the merging black holes turns into gravitational waves. 
The Big Bang mechanism in such a model is based on the Schwarzschild metric with variable mass 
\cite{Kutschera, GorkavyiA}.
The era of multiple merger of black holes begins when the Universe shrinks to a volume 
comparable to the total volume of existing BHs. This process should be natural for 
many cyclic models of the Universe \cite{Gurzadyan, Poplawski, GorkavyiD, GorkavyiB, GorkavyiC, Penrose}. 
It is easy to estimate that the total volume of 
BHs in the Universe is determined by the most massive SMBHs: if we take the average SMBH 
mass of $10^{8} M_\odot$ \cite{Nguyen}, then  $10^{11}$ such holes can be packed into a box with a size of 0.3 ly. 
If the SMBH mass is $10^{9} M_\odot$ \cite{Nguyen}, then even $10^{9}$ such holes will require a box with a size 
of 0.7 ly. Thus, when the size of the contracting Universe decreases by $\sim10^{10}$ times - up to 
several light years, the black holes will begin to merge en masse due to lack of space. 
There is an equilibrium value  $z_0$ that allows oscillations of the Universe, which does 
not accumulate heavy elements and is consistent with the observed population of black holes. 
If $z<z_0$ then the temperature of the Universe will be insufficient for 
photodissociation of iron nuclei. If $z>z_0$ then the total volume of observed 
black holes will be too large for such a small volume of the Universe. 
In \cite{GorkavyiC} the estimate $z_0\sim10^{10}$ was obtained independently of the data of NANOGrav. 

We hypothesize that the long-wavelength gravitational radiation recorded by NANOGrav 
\cite{Arz} is the relic gravitational radiation from the multiple merger of black holes 
during the Big Crunch: $f_{GWB}\sim{z_0}^{-1}f_{LIGO}$. Consequently, GWB is a fossil 
version of GW discovered by LIGO and stretched due to the expansion of the Universe. 
Let us show that the NANOGrav data can be explained within the framework of a cyclic model 
of the Universe with dark matter from black holes without putting forward new hypotheses or 
fitting free parameters. 

It is also assumed by \cite{Kohri, Vask} that the GWB detected by NANOGrav are generated 
during the formation of SBHs in the early stages of the Universe. In our 
model these SBHs come from the past cycles of the Universe. As will be shown below, our model 
and the hypothesis published by \cite{Kohri, Vask} have important differences in observational predictions.

\begin{figure}
    \includegraphics[width=\columnwidth]{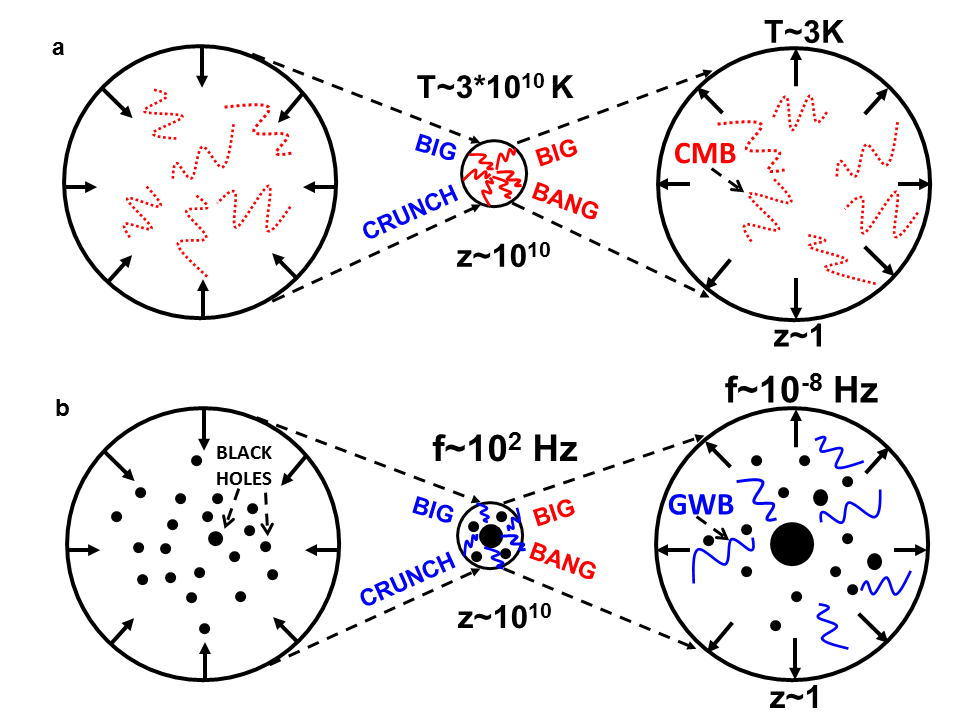}
    \caption{(From \cite{GorkavyiC}, with modifications). Evolution of relict electromagnetic waves (CMB) (a) 
	and gravitational radiation (GWB) (b) in the Universe during Big Crunch and Big Bang.}
    \label{fig:fg1} 
\end{figure}

\section{GWB in a one-component model} 
Let us evaluate the modern frequencies and amplitudes of GWs, which were formed 
during Big Crunch. The equation for the amplitude of gravitational radiation 
from a binary system of black holes of the same mass M, located at a distance 
D from the observer, can be written as \cite{Jaffe}:
\begin{equation}
h=4\sqrt{\frac{2}{5}}{({\frac{GM}{c^3}})^{5/3}}{({\frac{2\pi}{T}})^{2/3}}{\frac{c}{D}}, 
\label{eq:eq1}
\end{equation}
where $T$ is the proper rotation period of the binary system. 
Formula~(\ref{eq:eq1}) and subsequent formulas are obtained in the case of asymptotically 
flat spacetime. In the case of a closed Universe, the formulas for length and mass include 
the factor $(1-(D/R_U)^2)^{-1/2}$, where $R_U$ is the radius of 
curvature of the Universe (see, for example, \cite{Peebles}). We will use formula~(\ref{eq:eq1}) 
and subsequent formulas, assuming that the discussed distances are less than the 
radius of curvature of the Universe $(D/R_U)^2<<1$. The characteristic time of 
gravitational radiation of such a system can be obtained from the equation \cite{Jaffe}:
\begin{equation}
\frac{1}{\tau}={\frac{96}{5}}{({\frac{GM}{c^3}})^{5/3}}{({\frac{2\pi}{T}})^{8/3}}
\label{eq:eq2}
\end{equation}
The amplitude and frequency of the gravitational wave reach their maximum value and 
the period $T$ minimum value just before the merger of black holes. For this case, 
equations~(\ref{eq:eq1}) and~(\ref{eq:eq2}) are simplified. The maximum frequency 
of gravitational radiation when two identical black holes merge is written in the 
form (see, for example, \cite{Wen}):
\begin{equation}
f={\frac{2}{T(1+z)}}={\frac{c}{12\sqrt6\pi R(1+z)}}\approx{\frac{2.2*10^3}{{M_s}(1+z)}}
\label{eq:eq3}
\end{equation}
where $M_s$ is the mass of the black hole in solar masses, and $f$ is measured in hertz. 
The value $(1 + z)$ in the denominator describes the decrease in frequency due to the 
cosmological redshift, and the coefficient 2 in the numerator appears due to the 
quadrupole nature of gravitational radiation \cite{Arz}. From ~(\ref{eq:eq3}) we obtain that 
the merger of black holes with a mass of $\sim 10M_\odot$ (just such events were registered by 
LIGO \cite{Abbot16}) gives the natural frequency of the generated gravitational waves 
$f_0={2/T\approx 2.2*10^2}$Hz. 
Taking into account ~(\ref{eq:eq3}) and $z\sim1.6*10^{10}$ , we obtain the observed frequency of the gravitational 
stochastic radiation $f\sim1.4*10^{-8}$Hz, which coincides with the frequency of the 
waves discovered by NANOGrav. We obtain from (\ref{eq:eq3}) the minimum value of $T$:
\begin{equation}
T={\frac{24\sqrt6\pi R}{c}}.
\label{eq:eq4} 
\end{equation}
and substitute ~(\ref{eq:eq4}) in ~(\ref{eq:eq1}), taking into account the equation for the 
radius of the black hole $R={2GM}/{c^2}$. We get:
\begin{equation}
h=\frac{1}{3}\sqrt{\frac{2}{5}}{\frac{R}{D}}, 
\label{eq:eq5}
\end{equation}
For merging two BH with $R\approx10^7$cm and a distance $D\sim10^9$ ly, we get typical 
amplitude of GW, registered by LIGO: $h\sim10^{-21}$ \cite{Abbot16}. Equation (\ref{eq:eq2}) 
is also simplified for the case of maximum frequency (\ref{eq:eq3}):
\begin{equation}
\frac{1}{\tau}={\frac{4}{135}}{\frac{c}{R}}
\label{eq:eq6}
\end{equation}
To estimate the total energy of gravitational radiation from $N$ merging black holes, 
we will square the amplitude ~(\ref{eq:eq5}) and write:
\begin{equation}
h^2={\frac{2}{45}}{({\frac{R}{D}})^{2}}N{\frac{\tau}{t_0}}, 
\label{eq:eq7}
\end{equation}
where $t_0$ is the free path time of a BH moving at a speed $V$, before 
collision and merging with another BH \cite{Frolov}:  
\begin{equation}
{\frac{1}{t_0}}=108\pi R^2{nV}.
\label{eq:eq8}
\end{equation}
where $n={3N}/{(4\pi a^2)}$ is the concentration of black holes located in 
a sphere with radius $a$. The ratio ${\tau}/{t_0}$ characterizes the fraction 
of time during which a hole emits, merging with another hole. From 
Eq.~(\ref{eq:eq7}), taking into account Eq.~(\ref{eq:eq6}) and Eq.~(\ref{eq:eq8}), 
we obtain for the average amplitude of background gravitational waves 
generated by SBH during Big Crunch:
\begin{equation}
h\approx9\sqrt{\frac{3}{2}}{\frac{N R^{5/2}}{D a^{3/2}}}\sqrt{\frac{V}{c}}, 
\label{eq:eq9}
\end{equation}
where for the estimation it was taken: $N=10^{22}$ \cite{GorkavyiD}; $R=10^7 cm$;
$D\approx10^{10}$ ly; $a\approx3$ ly; $V\approx c$.
We assumed that the relative velocities of black holes in a maximally compressed 
Universe are comparable to the speed of light. The hypothesis about the 
accumulation of black holes and GWB from cycle to cycle of the Universe 
\cite{GorkavyiD} will change the meaning of the parameters in Eq.(\ref{eq:eq9}), 
for example, $N$ will mean the sum of GWB sources for a certain number of cycles.
Let's compare the amplitudes of the background gravitational waves from 
the SBH merger at Big Crunch $z\sim10^{10}$ and from the SMBH merger at a 
later time (at $z$ of the order of unity):
\begin{equation}
\frac{h_{SBH}}{h_{SMBH}}\approx({\frac{R_{SBH}}{R_{SMBH}}})^{5/2}({\frac{a_{SMBH}}{a_{SBH}}})^{3/2}{\frac{N_{SBH}}{N_{SMBH}}}A\sim10^{11}
\label{eq:eq10}
\end{equation}
The following estimates were accepted here: 
\begin{equation}
{\frac{R_{SBH}}{R_{SMBH}}}\sim10^{-6}; {\frac{a_{SMBH}}{a_{SBH}}}\sim10^{10}; {\frac{N_{SBH}}{N_{SMBH}}}\sim10^{11}
\label{eq:eq11}
\end{equation}
The last factor in~(\ref{eq:eq10}) is insignificant, because it is close 
enough to 1: 
\begin{equation}
A={\frac{D_{SMBH}}{D_{SBH}}}\sqrt{\frac{V_{SBH}}{V_{SMBH}}}\sim 1
\label{eq:eq12}
\end{equation}
We can obtain estimates of the background waves generated by black holes 
at different stages of the evolution of the Universe (see Table~\ref{Tab1}, where 
the mass of a SMBH is $10^6$ times the mass of an SBH).
\begin{table*}
\caption{Estimates of the amplitude and frequency of the GWB from different sources}
\label{Tab1}
\medskip
\begin{tabular}{|c|c|c|}
\hline
Source and z & Big Crunch, $z\sim10^{10}$  & $z\sim1$ \\
\hline
SBH & $h\sim10^{-15}$,$f\sim10^{-8} Hz$ & $h\sim10^{-30}$,$f\sim10^{2} Hz$ \\
SMBH & $h\sim10^{-11}$,$f\sim10^{-14} Hz$ & $h\sim10^{-26}$,$f\sim10^{-4} Hz$ \\
\hline
\end{tabular}
\end{table*}

It follows from (\ref{eq:eq10}) and Table~\ref{Tab1} that the amplitude of the stochastic 
gravitational waves caused 
by the SBH merger at the Big Crunch stage significantly exceeds the amplitude of the waves 
generated by modern SMBH mergers at the centers of galaxies. Note that estimates of emission 
from modern SMBHs usually consider not a moment of merger, but rather stable orbital of black 
holes around the center of mass \cite{Jaffe}. In this case, the radiation amplitudes of the individual 
system are less, but the radiation time $\tau$ in~(\ref{eq:eq7}) is longer. Unfortunately, there is a large 
uncertainty in such calculations, associated with the probability of merging galaxies and 
the problem of approaching the central SMBHs inside the last parsecs \cite{Jaffe,Wen,Goulding}.
Merging during Big Crunch SMBH with mass $\sim 10^7\div10^{10} M\odot$ will generate radiation with a 
frequency of $f_0\sim10^{-(4\div7)}$ Hz. Taking into account $z\sim10^{10}$Hz, we get that in the 
modern epoch there should be very long gravitational waves with a frequency  
$f\sim10^{-(14\div17)}$ Hz. Such waves have a length of $\sim10^{-(1\div4)}$ of the size 
of the Universe. It follows from ~(\ref{eq:eq9}) that the SMBH fusion will generate a GWB with 
a significant amplitude. For example, for a frequency $f\sim10^{-14}$ Hz, the GW 
amplitude will be $\sim10^{4}$ times greater than the amplitude of GW discovered by 
NANOGrav \cite{Arz} (see Table~\ref{Tab1}). The discovery of such wavelengths is possible using 
data from the Gaia astrometric satellite \cite{Gwinn, Book, Moore}.

\section{GWB in a multi-component model} 
In our estimates, we assumed that all BHs have the same mass. Let us assume that the mass 
distribution of SBHs participating in mergers during Big Crunch is close to the modern 
distribution of black holes because a population of relic black holes can form from 
astrophysical black holes that accumulate in the repeating cycles of the 
Universe \cite{GorkavyiD}. The observed population of black holes consists mainly of SBH 
from 4 to 100 $M\odot$ \cite{GorkavyiD,Abbot20} and SMBH with $\sim10^5-10^{10}M\odot$ 
located in the centers of galaxies \cite{Cherep}.

\begin{figure}
    \includegraphics[width=\columnwidth]{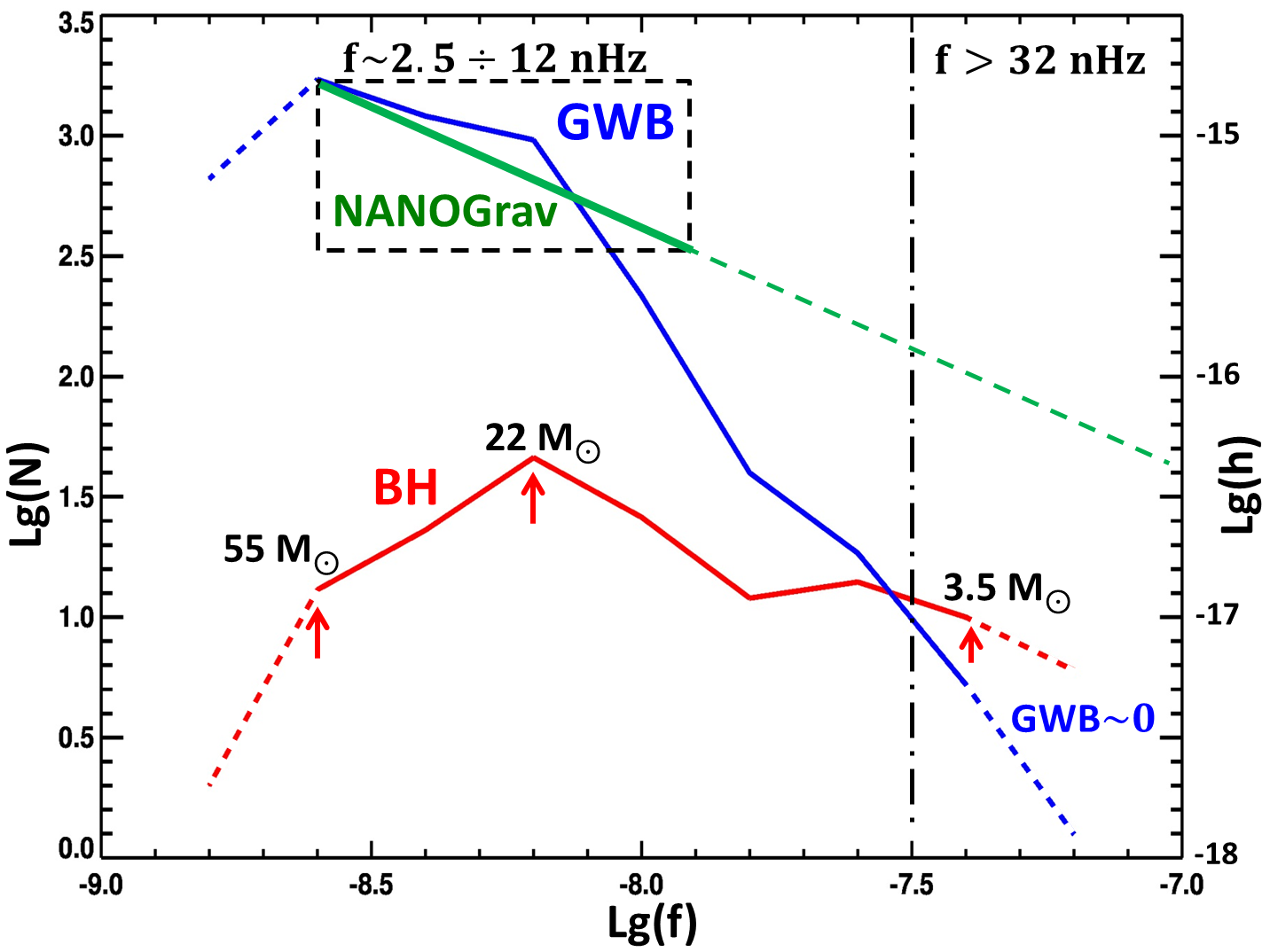}
    \caption{Solid curve BH: mass spectrum of SBH according to LIGO data with the 
	addition of black holes in binary stellar systems \cite{Abbot20,Cherep} as a function of GW 
	frequency (arrows indicate the magnitude of mass of BHs corresponding to the specific frequency 
	of GWB). Solid line GWB shows the theoretically calculated 
	amplitude of gravitational radiation during 
	the SBH merger. Dashed parts of curves BH and GWB show zones with low statistics. 
	The dashed rectangle shows the border corresponding to the frequency of 
	gravitational waves discovered by NANOGrav (thick straight line, the maximum 
	of which is aligned with the maximum of the model curve). The dash-dotted 
	vertical line shows the border corresponding to the frequency of gravitational 
	waves of $3.2*10^{-8} Hz$ (or 1/year) from the merging of black holes at 
	$\sim4M\odot$, so to the right of this line there will be a strong deficit of GWB 
	in comparison with the extrapolated NANOGrav data (dashed continuation of a thick 
	straight line) and theoretical models based on the gravitational waves from 
	SMBHs \cite{Jaffe}.}
    \label{fig:fg2} 
\end{figure}

It should be taken into account that formula~(\ref{eq:eq9}) includes the mass distribution of black 
holes $N(m)$. Figure~\ref{fig:fg2} shows the spectrum of mass of BH as a function of 
frequency $h(f)$ under the assumption that merging black holes generate 
gravitational radiation of one wavelength (with maximum frequency and 
amplitude).
Figure~\ref{fig:fg2} shows that the detected gravitational waves are located near the 
maximum amplitude of the gravitational waves, which follow from our model. 
 
Also Figure~\ref{fig:fg2} shows that due to the influence of the mass spectrum of black 
holes, the frequency dependence of the amplitude can vary significantly. 
For $f\approx (2.5\div6.3)*10^{-9}$Hz (or $2.5\div6.3$ nHz) our model gives $h\propto f^{-(0.5\div0.6)}$, 
which close to the observation of NANOGrav (see Figure~\ref{fig:fg2}). The spectrum of 
gravitational radiation from small ($4\div20 M\odot$) SBHs is dependent on 
observational selection, so it is difficult to draw any reliable 
conclusions for $f>6.3*10^{-9}$Hz (or $6.3$ nHz). The most reliable observational 
fact is the absence of black holes less than 4 solar masses 
$f>3.2*10^{-8}$Hz (or $f>32$ nHz), which should be reflected in the spectrum of gravitational 
waves. The $f = lg(3.2*10^{-8})=-7.5$ barrier (marked with a vertical line) 
corresponds to the period of gravitational waves in one year. According to 
observations there are no astrophysical black holes less than 4 times the 
mass of the Sun, or there are very few of them. This should cause a strong 
deficiency of GWB with frequencies $f>3.2*10^{-8}$Hz (see Figure~\ref{fig:fg2}). 
Indeed, no ''any detectable contributions from a GWB'' in frequency 
$f>1.2*10^{-8}$Hz (or $12$ nHz) \cite{Arz}. We consider this fact to be very important. 
The deficit (beyond the expected power-law dependences) of GWB at frequencies 
of $f>3.2*10^{-8}$Hz (or $f>32$ nHz) will confirm our model.

\section{Discussion} 
Stochastic gravitational waves must be accompanied by a quadrupole spatial correlation according to the 
Hellings-Downs curve \cite{Hellings}. As the results of NANOGrav show, the background of 
stochastic gravitational waves has no signs of monopole and dipole waves. At the same 
time, GWB has some signs of quadrupole spatial correlation, but with a low level of 
statistical confidence \cite{Arz}. After submitting this article to the journal, 
the article \cite{Goncharov} was published with the results of the analysis of the 
Parkes Pulsar Timing Array (PPTA). Under the assumptions of an analysis of the NANOGrav, 
the team of PPTA have detected with high confidence a common-spectrum time-correlated 
signal in the timing of the 26 PPTA-DR2 millisecond pulsars: $h\sim 2.2*10^{-15}$ at a 
frequency of 1 ${yr}^{-1}$ \cite{Goncharov}. The PPTA data also showed statistically weak 
indications of quadrupole spatial correlation. \cite{Romano} show that the current lack 
of evidence for quadrupolar spatial correlations is consistent with the magnitude of the 
correlation coefficients for pairs of Earth-pulsar baselines in the array, and the fact that 
pulsar timing arrays are most-likely operating in the intermediate-signal regime. We believe 
that the weak quadrupole correlation found in both works (Fig.7 \cite{Arz} and Fig.3 \cite{Goncharov}) 
will increase with the amount of data used. The discussions around the 
GWB (\cite{Arz}; \cite{Goncharov}; \cite{Romano}) show that the statistical methods used in 
the analysis of pulsar signals contain significant assumptions that must be examined to 
finally prove the validity of the discovery. From our point of view, the reliability of the 
discovery of the stochastic background of gravitational waves is confirmed by the fact that 
the existence of a powerful background of nanohertz relict waves was predicted in the 
article \cite{GorkavyiD}, submitted to the journal in May 2020.

\section{Conclusions}

LIGO detectors observe gravitational waves from rare modern mergers of SBHs - the most 
numerous black holes in the Universe. We believe that the telescopes of the NANOGrav and the PPTA 
detected a burst of similar gravitational radiation from the numerous SBH mergers 
during Big Crunch. Decrease in $10^{10}$ times the frequency of these relic gravitational 
waves is associated with the expansion of the Universe. We believe that the NANOGrav and the PPTA 
data support the hypothesis that black holes can make up the bulk of dark matter 
\cite{Bird,Kashlinsky,Clesse} and that a significant portion of the observed black holes came 
from the past cycle of the oscillating Universe \cite{Gurzadyan,Clifton}.
We find it remarkable that the parameter $z\sim10^{10}$, 
which satisfies the model of a cyclic Universe, turns out to be just such that 
the frequency of relic gravitational waves coincides with the frequency of 
gravitational waves observed by NANOGrav and PPTA. 

From our model, specific conclusions can be drawn that can be verified 
by observations:
 
1. The spectrum of GWB discovered by NANOGrav and PPTA  will not be described by 
Planck's law or any other law related to thermal equilibrium. The GWB should 
reflect the observed distribution of black holes (SBHs and SMBHs). 

2. The amplitude of GWB discovered by NANOGrav and PPTA decreases with increasing 
frequency, but the rate of this decrease will become especially noticeable 
at frequency $f>3.2*10^{-8}$Hz (or $f>32$ nHz) (the oscillation period is less 
than a year) due to a lack of astrophysical black holes with masses $<4M\odot$.
This feature of the spectrum cannot be predicted by any model based on the 
hypothesis of primordial black holes born in the Big Bang (see, for example,
\cite{Kohri}).

3. The numerous SMBHs that are currently observed in the centers of 
galaxies should generate low frequency GWB $f\sim10^{-(14\div17)}$Hz 
during the Big Crunch.

\section*{Declaration of Competing Interest}

The author declares that he has no known competing financial interests or personal relationships 
that could have appeared to influence the work reported in this paper.

\section*{Acknowledgments}

The author thanks John Mather, Alexander Vasilkov, Sergei Kopeikin and George Hobbs 
for helpful discussions and comments.






\bibliography{mybibfile45}

\end{document}